\documentclass[11pt]{article}
\usepackage{moriond,epsfig}

\def\Journal#1#2#3#4{{#1} {\bf #2}, #3 (#4)}

\def\PLB{{\em Phys. Lett.}  B}
\def\PRL{\em Phys. Rev. Lett.}
\def\PRD{{\em Phys. Rev.} D}

\def\be{\begin{equation}}
\def\ee{\end{equation}}
\def\bea{\begin{eqnarray}}
\def\eea{\end{eqnarray}}

\RequirePackage{xspace}
\usepackage{relsize}
\def\babar{\mbox{\slshape B\kern-0.1em{\smaller A}\kern-0.1em
    B\kern-0.1em{\smaller A\kern-0.2em R}}}
    
\def\en         {\ensuremath{e^-}\xspace}   % electron negative (\em is taken)
\def\ep         {\ensuremath{e^+}\xspace}

\def\u     {\ensuremath{u}\xspace}
\def\ubar  {\ensuremath{\overline u}\xspace}

\def\d     {\ensuremath{d}\xspace}
\def\dbar  {\ensuremath{\overline d}\xspace}

\def\s     {\ensuremath{s}\xspace}
\def\sbar  {\ensuremath{\overline s}\xspace}

\def\c     {\ensuremath{c}\xspace}
\def\cbar  {\ensuremath{\overline c}\xspace}

\def\b     {\ensuremath{b}\xspace}

\def\piz   {\ensuremath{\pi^0}\xspace}

\def\pip   {\ensuremath{\pi^+}\xspace}
\def\pim   {\ensuremath{\pi^-}\xspace}

\def\Kbar  {\kern 0.2em\overline{\kern -0.2em K}{}\xspace}

\def\Kz    {\ensuremath{K^0}\xspace}
\def\Kzb   {\ensuremath{\Kbar^0}\xspace}
\def\KzKzb {\ensuremath{\Kz \kern -0.16em \Kzb}\xspace}
\def\Kp    {\ensuremath{K^+}\xspace}
\def\Km    {\ensuremath{K^-}\xspace}

\def\KpKm  {\ensuremath{\Kp \kern -0.16em \Km}\xspace}
\def\KS    {\ensuremath{K^0_{\scriptscriptstyle S}}\xspace}

\def\Kstarp  {\ensuremath{K^{*+}}\xspace}

\def\Dbar    {\kern 0.2em\overline{\kern -0.2em D}{}\xspace}
\def\Db      {\ensuremath{\Dbar}\xspace}
\def\D      {\ensuremath{D}\xspace}
\def\Dz      {\ensuremath{D^0}\xspace}
\def\Dzb     {\ensuremath{\Dbar^0}\xspace}
\def\DzDzb   {\ensuremath{\Dz {\kern -0.16em \Dzb}}\xspace}
\def\Dp      {\ensuremath{D^+}\xspace}
\def\Dm      {\ensuremath{D^-}\xspace}

\def\DpDm    {\ensuremath{\Dp {\kern -0.16em \Dm}}\xspace}

\def\Dstarp  {\ensuremath{D^{*+}}\xspace}
\def\Dstarm  {\ensuremath{D^{*-}}\xspace}

\def\Dsp      {\ensuremath{D_s}\xspace}
\def\Dsp      {\ensuremath{D^+_s}\xspace}

\def\X       {\ensuremath{X}\xspace}

\def\B       {\ensuremath{B}\xspace}
\def\Bbar    {\kern 0.18em\overline{\kern -0.18em B}{}\xspace}
\def\Bb      {\ensuremath{\Bbar}\xspace}

\def\Bz      {\ensuremath{B^0}\xspace}
\def\Bzb     {\ensuremath{\Bbar^0}\xspace}
\def\BzBzb   {\ensuremath{\Bz {\kern -0.16em \Bzb}}\xspace}
\def\Bu      {\ensuremath{B^+}\xspace}
\def\Bub     {\ensuremath{B^-}\xspace}
\def\Bp      {\ensuremath{\Bu}\xspace}
\def\Bm      {\ensuremath{\Bub}\xspace}

\def\BpBm    {\ensuremath{\Bu {\kern -0.16em \Bub}}\xspace}

\def\BorBbar    {\kern 0.18em\optbar{\kern -0.18em B}{}\xspace}
\def\DorDbar    {\kern 0.18em\optbar{\kern -0.18em D}{}\xspace}
\def\KorKbar    {\kern 0.18em\optbar{\kern -0.18em K}{}\xspace}

\def\jpsi     {\ensuremath{{J\mskip -3mu/\mskip -2mu\psi\mskip 2mu}}\xspace}

\mathchardef\Upsilon="7107
\def\Y#1S{\ensuremath{\Upsilon{(#1S)}}\xspace}% no space before {...}!

\mathchardef\Deltares="7101
\mathchardef\Xi="7104
\mathchardef\Lambda="7103
\mathchardef\Sigma="7106
\mathchardef\Omega="710A

\def\Deltabar{\kern 0.25em\overline{\kern -0.25em \Deltares}{}\xspace}
\def\Lbar{\kern 0.2em\overline{\kern -0.2em\Lambda\kern 0.05em}\kern-0.05em{}\xspace}
\def\Sigbar{\kern 0.2em\overline{\kern -0.2em \Sigma}{}\xspace}
\def\Xibar{\kern 0.2em\overline{\kern -0.2em \Xi}{}\xspace}
\def\Obar{\kern 0.2em\overline{\kern -0.2em \Omega}{}\xspace}
\def\Nbar{\kern 0.2em\overline{\kern -0.2em N}{}\xspace}
\def\Xb{\kern 0.2em\overline{\kern -0.2em X}{}\xspace}

\def\X {\ensuremath{X}\xspace}

\def\BR         {{\ensuremath{\cal B}\xspace}}

\newcommand{\tev}{\ensuremath{\mathrm{\,Te\kern -0.1em V}}\xspace}
\newcommand{\gev}{\ensuremath{\mathrm{\,Ge\kern -0.1em V}}\xspace}
\newcommand{\mev}{\ensuremath{\mathrm{\,Me\kern -0.1em V}}\xspace}
\newcommand{\kev}{\ensuremath{\mathrm{\,ke\kern -0.1em V}}\xspace}
\newcommand{\ev}{\ensuremath{\mathrm{\,e\kern -0.1em V}}\xspace}
\newcommand{\gevc}{\ensuremath{{\mathrm{\,Ge\kern -0.1em V\!/}c}}\xspace}
\newcommand{\mevc}{\ensuremath{{\mathrm{\,Me\kern -0.1em V\!/}c}}\xspace}
\newcommand{\gevcc}{\ensuremath{{\mathrm{\,Ge\kern -0.1em V\!/}c^2}}\xspace}
\newcommand{\mevcc}{\ensuremath{{\mathrm{\,Me\kern -0.1em V\!/}c^2}}\xspace}
\def\fb   {\ensuremath{\mbox{\,fb}}\xspace}
\def\invfb   {\ensuremath{\mbox{\,fb}^{-1}}\xspace}

\def\mus  {\ensuremath{\rm \,\mus}\xspace}

\def\mus        {\ensuremath{\,\mu{\rm s}}\xspace}    %% microsecond
\def\to                 {\ensuremath{\rightarrow}\xspace}

\def\pep2{PEP-II}

\def\gsim{{~\raise.15em\hbox{$>$}\kern-.85em
          \lower.35em\hbox{$\sim$}~}\xspace}
\def\lsim{{~\raise.15em\hbox{$<$}\kern-.85em
          \lower.35em\hbox{$\sim$}~}\xspace}

\def\CP                {\ensuremath{C\!P}\xspace}
\def\Vud  {\ensuremath{V_{ud}}\xspace}
\def\Vcd  {\ensuremath{V_{cd}}\xspace}
\def\Vtd  {\ensuremath{V_{td}}\xspace}

\def\Vub  {\ensuremath{V_{ub}}\xspace}
\def\Vcb  {\ensuremath{V_{cb}}\xspace}
\def\Vtb  {\ensuremath{V_{tb}}\xspace}
\xspace

\def\beq {\begin{equation}}
\def\eeq {\end{equation}}

\begin{document}
\begin{flushleft}
SLAC-PUB-11271\\
\babar-PROC-05-019
\end{flushleft}

\vspace*{2.5cm}
\title{HADRONIC \B AND $D$ STUDIES AT \babar}

\author{F. COUDERC\\representing the \babar\ Collaboration}

\address{Laboratoire d'Annecy-Le-Vieux de Physique des Particules\\ 9 Chemin de Bellevue\\
         BP 110 F-74941 Annecy-le-Vieux CEDEX - France}

\maketitle\abstracts{
We present new results on hadronic \B and $D$ decays from the \babar\ experiment. The first part of this document presents searches for new channels which may be used for \CP\ measurements. The second part is dedicated to hadronic decays with tests of QCD factorization predictions and other models for \B structure and decay mechanisms. A new result on the reference branching ratio $\D_s^+\to\phi\pip$ is also reported. }

\section{\CP related analysis}

In this section, we present  new \CP\ violation related analysis of channels which could be used, in the future, to measure some of the CKM matrix parameters like $\gamma\equiv \mathrm{arg}\left[-\frac{\Vud\Vub^*}{\Vcd\Vcb^*}\right]$ or a combination of $\gamma$ and $\beta\equiv\mathrm{arg}\left[-\frac{\Vcd\Vcb^*}{\Vtd\Vtb^*}\right]$. To do this, one needs to measure \CP\ asymmetries which are defined, for the $\B\to f$ decay, as~:
    \begin{displaymath}
    \mathcal{A}_{\CP} = \frac{\BR(\B\to f)-\BR(\Bb\to\overline{f})}{\BR(\B\to f)+\BR(\Bb\to\overline{f})}
    \end{displaymath}
    
\subsection{Measurement of the branching fraction and decay rate asymmetry of $\Bm\to D_{\pip\pim\piz}\Km$}

The decays $\B\to D^{(*)0} K^{(*)}$ can be used to measure the angle $\gamma$ taking advantage of the interference between $\b\to\u\cbar\s$ and $\b\to\c\ubar\s$ decay amplitudes. Different approaches have been developed, and among which, $\gamma$ measurements involving \D decays to multi-body, using a Dalitz plot analysis technique as described in reference~\cite{cite:DK_interfer}. In this analysis, we study the decay mode $\Bm\to\D\Km$ with the \D-decay~: $\D\to\pip\pim\piz$ which is Cabibbo suppressed. This yields a much smaller event sample compared to Cabibbo allowed decay but its interfering \Dz and \Dzb amplitudes have similar magnitudes. Due to these interferences, the production rate might be different from the product $\BR_{prod}\equiv \BR(\Bm\to\Dz\Km)\times\BR(\Dz\to\pip\pim\piz)= (4.1\pm1.6)\times 10^{-6}$ (ref.~\cite{cite:PDG}). From a sample of 229 million of $\B\Bb$ pairs, we found $133\pm23$ signal events which correspond to a branching ratio of 
    $\BR(\Bm\to D_{\pip\pim\piz}\Km) = (5.5\pm1.0\pm0.7)\times 10^{-6}$. We determine the raw asymmetry and do not find any significant deviation from zero~: 
    $\mathcal{A}_{\CP}^{raw} = 0.02\pm0.16\pm0.03$.  
The $\gamma$ extraction needs a full Dalitz analysis of the \D-decay.

\subsection{Search for $\B\to\Dsp\X_{light}$ with $\X_{light}\equiv\piz,\, a_0^-,\, a_1^-$}

The value of $\sin(2\beta+\gamma)$ can be extracted from the measurement of the time dependent \CP asymmetry in $\B\to\Dm\X^{+}_{light}$ decays with, for instance, $\X^+_{light}\equiv\pip,\, a_0^+,\, a_2^+$. In this case the asymmetry is given by~: $\mathcal{A}_{\CP}(\Delta t) = r\times\sin(2\beta+\gamma)\times sin(\Delta m_{d}\Delta t)$ where $r=\BR(\Bz\to\Dp\X^-_{light})/\BR(\Bz\to\Dm\X^+_{light})$. The decay $\Bz\to\Dp\X^-_{light}$ is doubly Cabibbo suppressed and difficult to measure directly. Using $SU(3)$ flavor symmetry, it is possible to infer the value of $\BR(\Bz\to\Dp\X^-_{light})$ from the value of $\BR(\B\to\Dsp\X_{light})$, the latter being less suppressed.\\
If $\X_{light}^+\equiv\pip$, then $r$ is expected to be very small ($r\approx0.02$) which implies a small asymmetry. In this case $r$ might be deduced from the rate $\BR(\Bp\to\Dsp\piz)$. We measure this branching ratio from a sample of $124$ millions of $\B\Bb$ pair, we do not see any significant signal and quote an upper limit of~:
 $\BR(\Bp\to\Dsp\piz)<2.8\times10^{-5}$
in agreement with a previous measurement by CLEO ($<2.4\times10^{-4}$ from ref.~\cite{cite:PDG}) and with the value of $0.9\times 10^{-5}$ expected from the rate of $\BR(\Bz\to\Dsp\pim)$ measured by Belle and \babar\ experiments. If $\X_{light}^+\equiv a_0^+\, (a_2^+)$, $r$ might be quite large. This is due to the coupling constant of the $W$ to the $a_0$ scalar meson ($a_{2}$ meson of spin 2) which is small and decreases the production rate of the Cabibbo allowed decay $\Bz\to\Dm a_0^+(a_2^+)$. The factorization hypothesis predicts a similar rate for Cabibbo allowed and Cabibbo suppressed decays~\cite{cite:Da_facto} which results in $r\approx 1$. These decays are not yet in the reach of experiment (branching ratios around $10^{-6}$), nevertheless, the theoretical predictions can be tested with the measurement of the branching ratio of the decay $\Bz\to\Dsp a_0^- (a_2^-)$ expected at larger values~: $\BR(\Bz\to\Dsp a_0^-(a_2^-)\approx 7.5(1.5)\times 10^{-5}$ (ref.~\cite{cite:Da_facto,cite:Dsa_facto}). From a sample of $230$ million of $\B\Bb$ pairs, we measure these two branching ratios. The $a_0^-\, (a_2^-)$ is reconstructed in $a_{0,2}^-\to\eta(\to\gamma\gamma)\pip$ which has a branching ratio of the order of 100~\% (only 15~\% for the $a_2^-$ which lowers the experimental sensitivity). We do not find any significant signal and quote the upper limits~:
    $
    \BR(\Bz\to\Dsp a_{0(2)}^-) < 4.0\,(25)\,\times 10^{-5}
    $
which shows a discrepancy of at least a factor two with the theoretical predictions.

\subsection{Charmless decays}

The decay $\Bp\to\Kstarp(\to\Kp\piz)\piz$ and its \CP\ asymmetry is particularly interesting in light of the recent measurement of direct \CP\ violation in the decay $\Bz\to\Kp\pim$~\cite{cite:Kpi_viol}. From a sample of $232$ million of $\B\Bb$ pairs we find $88.5\pm25.7$ signal events which correspond to the branching ratio~: $\BR(\Bp\to\Kstarp\piz) = (6.9\pm2.0\pm1.3)\times10^{-6}$ and do not find any hint of direct \CP\ violation~: $\mathcal{A}_{\CP} = 0.04\pm0.29\pm0.05$ (reference~\cite{cite:Kpi_babar}).\\
Other types of charmless decays such as~: $\B\to\eta^{(')}\pi,\,\rho,\,\omega,\, K$ are studied as well. These decays process via $\b\to\u$ CKM-suppressed trees and/or $\b\to\s$ penguins. They provide a good test of QCD factorization, may have large \CP\ asymmetry and may provide sensitivity to the CKM angle $\gamma$. From a sample of $232$ million of $\B\Bb$ pairs, we measure the four branching ratios, and their asymmetries, and give two upper limits. Details on this analysis can be found in reference~\cite{cite:EtaX_babar}, no significant \CP\ asymmetry is observed and measured branching ratios are in good agreement with theoretical predictions. We make the first observation of the decays $\Bp\to\eta^{'}\pip$ and $\Bp\to\eta\rho^+$ and measure the corresponding decay rates~: $\BR(\Bp\to\eta^{'}\pip)=(4.0\pm0.8\pm0.4)\times10^{-6}$ and $\BR(\Bp\to\eta\rho^+)=(8.4\pm1.9\pm1.1)\times10^{-6}$.

\section{Non \CP related analysis}

\subsection{Measurement of the $\Bz\to\Dstarm D_s^{*+}$ and $D_s^+\to\phi\pip$ branching ratios}

We present measurements of the branching ratio $\BR(\Bz\to\Dstarm D_s^{*+})$ which may lead to a precise determination of the reference $\BR(D_s^+\to\phi\pi^+)$. They have been performed on a sample of $123$ million of $\B\Bb$ pairs. The $\Bz\to\Dstarm D_s^{*+}\to(\Dzb\pim)(D_s^+\gamma)$ decay is reconstructed using two different methods. The first one combines the fully reconstructed $\Dstarm$ with the photon from the $D_s^{*+}\to D_s^+\gamma$ decay, without explicit reconstruction of the $D_s^+$. To extract the number of partially reconstructed events, we compute the "missing mass" $m_{miss}$ recoiling against the $\Dstarm\gamma$ system assuming that a $\Bz\to\Dstarm D_s^{*+}\to(\Dzb\pim)(D_s^+\gamma)$ decay took place. For signal events, $m_{miss}$ peaks at the $D_s$ mass. We find, with this method, the following branching ratio~:
    $
    \BR_1 \equiv \BR(\Bz\to\Dstarm D_s^{*+})
          = (1.88\pm0.09\pm0.17)~\%
    $
which is in agreement with the factorization model~: $\BR(\Bz\to\Dstarm D_s^{*+})_{theo} = (2.4\pm 0.7)~\%$. The second one is a full reconstruction technique of the decay chain $\Bz\to\Dstarm\D_s^{*+}$ where the $\D_s$  candidate is reconstructed in the mode~: $D_s^+\to\phi\pip\to(\Kp\Km)\pip$. We measure with this method, the branching ratio 
    $\BR_2\equiv \BR(\Bz\to\Dstarm D_s^{*+})\times\BR(D_s^+\to\phi\pip) 
    = (8.81\pm0.86_{stat})\times 10^{-4}$. \\
From the ratio $\BR_2/\BR_1$, where many systematics cancel out, we get a precise measurement of~:
    \begin{displaymath}
    \BR(D_s^+\to\phi\pip) = (4.81\pm0.52\pm0.38)~\%.
    \end{displaymath}
which represents an improvement on the error by about a factor of two over previous measurements~\cite{cite:PDG}. Details on this work can be found in the reference~\cite{cite:hep_DsD}.

\subsection{Search for the rare decays $\Bp\to D^{(*)+}\Kz$}

This decay is expected to occur via a pure annihilation diagram. Such processes provide interesting insights into the internal dynamics of \B mesons. This kind of diagram cannot be calculated in QCD factorization since both the quarks play a role. These amplitudes are expected to be suppressed, with respect to the amplitudes of spectator quark trees, by a factor $f_{\B}/m_{\B}\approx 0.04$ and have never been observed. Some studies~\cite{cite:DpKs_rescattering} indicate , though, that processes with a spectator quark can contribute to annihilation-mediated decays by \emph{rescattering}. The branching ratio is expected to be either at the level of the current sensitivity ($10^{-5}$) if large rescattering occurs, or three orders of magnitude below if not~\cite{cite:DpKs_rescattering}. We reconstruct the two decay modes $\Bp\to\Dstarp\KS$ and $\Bp\to\Dp\KS$ within a sample of $226$ million of $\B\Bb$ pairs. We do not see any significant excess of signal and therefore we set the upper limits at 90~\% CL~: $\BR(\Bp\to\Dp\KS)<0.5\times10^{-5}$ and $\BR(\Bp\to\Dstarp\KS)<0.9\times10^{-5}$. 

\subsection{Search for $\B\to\jpsi D$ Decays}

The spectra of the momentum of inclusive \jpsi mesons in the \Y4S rest frame observed by CLEO and by \babar, compared with calculations using non-relativistic QCD (NRQCD), show an excess at low momentum, corresponding to a branching fraction of approximately $6\times10^{-4}$. Many hypothesis have been proposed to explain this result but no experimental evidence has been found to support them. The presence of $\b\ubar\c\cbar$ components (intrinsic charm) in the \B-meson wave function has also been suggested to enhance the branching ratio of decays such as $\B\to\jpsi \Db(\pi)$ to the order of $10^{-4}$ while pertubative QCD predicts a branching ratio for $\B\to\jpsi \Db$ of $10^{-8}$-$10^{-9}$. We test the decay channels $\B\to\jpsi D$ within a sample of $124$ million of $\B\Bb$ pairs. We do not find any evidence of signal and obtain upper limits of $1.3\times10^{-5}$ for $\Bz\to\jpsi\Dzb$ and $1.2\times 10^{-4}$ for $\Bp\to\jpsi\Dp$ at 90~\% CL. Therefore, we conclude that intrinsic charm is not supported as the explanation of low momentum \jpsi excess in \B decays. More details on this analysis can be found in reference~\cite{cite:jpsiD_hep}.

\subsection{Production and decay of the $\Xi_c^0$ at \babar}

\begin{figure}[!h]
    \begin{center}
    \includegraphics[width=10cm]{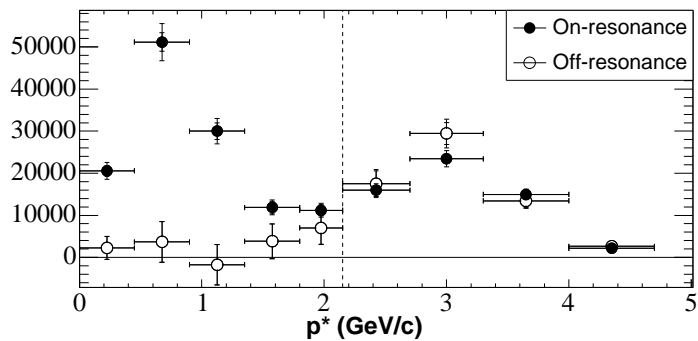}
    \caption{The $p^*$ spectrum of $\Xi_c^0$. Black dots represent on-peak data. Empty dots are off-resonance data ($\ep\en\to\u\ubar/\d\dbar/\s\sbar/\c\cbar$ only) scaled to account for the difference in integrated luminosity and cross-section with on-peak data.}
    \end{center}
    \label{fig:pstar_Xic}
\end{figure}

We present a study of the $\Xi_c^0~(\c\s\d)$ charmed baryon using a luminosity of $116.1~\invfb$ through two decay modes~: $\Xi_c^0\to\Omega^-\Kp$ and $\Xi_c^0\to\Xi^-\pip$. First, we measure the ratio of these two decay rates to be $0.294\pm0.018\pm0.016$ which is compatible with the prediction, in a spectator quark model calculation, of 0.32. Then, we measure the $p^*$ distribution of the $\Xi_c^0$ baryons, in the \Y4S frame, in order to study the production mechanisms in both $\c\cbar$ and $\B\Bb$ events. Results are shown on Figure~\ref{fig:pstar_Xic}. The double-peak structure seen in the $p^*$ spectrum is due to two production mechanisms: the peak at lower $p^*$ is due to $\Xi_c^0$ production in \B meson decays and the peak at higher $p^*$ is due to $\Xi_c^0$ production from the $\c\cbar$ continuum. From these spectra we compute the cross-section of the production of $\Xi_c^0$ in continuum~: $\sigma(\ep\en\to\c\cbar\to\Xi_c^0\X)\times\BR(\Xi_c^0\to\Xi^-\pip) = (388\pm39\pm41)\fb$ and the rate of $\Xi_c^0$ production in \B decay~:
$\BR(\B\to\Xi_c^0\X)\times\BR(\Xi_c^0\to\Xi^-\pip) = (2.11\pm0.19\pm0.25)\times10^{-4}$.\\

The high rate of $\Xi_c^0$ production at low $p^*$ in \B decays (below $1.2~\gevc$) is particularly intriguing, it implies that the invariant mass of the recoiling antibaryon system is typically above $2.0~\gevcc$.
This can be explained naturally by a substantial rate of charmed baryon pair production through the $\b\to\c\cbar\s$ weak decay process which was observed indirectly in a previous \babar\ analysis. Further details on this $\Xi_c^0$ production analysis can be found in reference~\cite{cite:Xic_paper}.


\section*{References}
\begin{thebibliography}{99}
%\bibitem{cite:DK_idea} M. Gronau and D. Wyler, \Journal{\PLB}{265}{172}{1991}.
\bibitem{cite:DK_interfer} A. Giri, Y. Grossman, A. Soffer and J. Zupan, \Journal{\PRD}{68}{054018}{2003}.
%\bibitem{cite:DspPim_CLEO}
%\bibitem{cite:DspPim_Bfactories} B. Aubert et al. [\babar\ collaboration], \Journal{\PRL}{90}{181803}{2003};
%                                 K. Abe et al. [Belle collaboration], \verb|hep-ex|/0408109.
\bibitem{cite:PDG} S. Eidelman {\it et al.} [Particle Data Group], \Journal{\PLB}{592}{1}{2004}.
\bibitem{cite:Da_facto} M. Diehl, G. Hiller, \verb|hep-ph|/0105194
\bibitem{cite:Dsa_facto} C.S. Kim, J.P. Lee, and M. Bauer, \Journal{Z. Phys.~\textbf{C}}{29}{637}{1985}.
\bibitem{cite:Kpi_viol} B. Aubert {\it et  al.} [\babar\ collaboration], \Journal{\PRL}{93}{131801}{2004};
                        Y. Chao {\it et al.} [Belle collaboration], \Journal{\PRL}{93}{191802}{2004}.
\bibitem{cite:Kpi_babar} B. Aubert {\it et al.} [\babar\ collaboration], \verb|hep-ex|/0504009.
\bibitem{cite:EtaX_babar} B. Aubert {\it et al.} [\babar\ collaboration], \verb|hep-ex|/0408058,
                                                                            \verb|hep-ex|/0503035.
%%%%%%%%%%%% Ds D
%\bibitem{cite:factorization_DsD} M. Beneke, \Journal{J. Phys.~\textbf{G}}{27}{1069}{2001}, and references therein.
%\bibitem{cite:DsD_theo} Z. Luo and J.L. Rosner, \Journal{\PRD}{64}{094001}{2001}.
%\bibitem{cite:DsD_CLEO} M.Artuso {\it et al.} [CLEO collaboration], \Journal{\PLB}{378}{364}{1996}.
\bibitem{cite:hep_DsD} B. Aubert {\it et al.} [\babar\ collaboration], \verb|hep-ex|/0502041 and references therein.

%%%%%%%%%%%%%% D+ Ks
\bibitem{cite:DpKs_rescattering} M. Blok, M. Gronau and J.L. Rosner, \Journal{\PRL}{78}{3999}{1997}.

%%%%%% Jpsi D
%\bibitem{cite:jpsi_CLEO} R. Balest {\it et al.} [CLEO collaboration], \Journal{\PRD}{52}{2661}{1995}.
%\bibitem{cite:jpsi_babar} B. Aubert {\it et al.} [\babar\ collaboration], \Journal{\PRD}{67}{032002}{2003}.
%\bibitem{cite:jspi_NRQCD} M. Beneke, G.A. Schuler and S. Wolf, \Journal{\PRD}{62}{034004}{2000}.
%\bibitem{cite:jpsi_theo1} S.J. Brodsky and F.S. Navarra, \Journal{\PLB}{411}{152}{1997}.
%\bibitem{cite:jpsi_theo2} C.T.H. Davies and S.H.H. Tye, \Journal{\PLB}{154}{332}{1985}; C.K. Chua, W.S. Hou and G.G. Wong, \Journal{\PRD}{68}{054012}{2003}.
%\bibitem{cite:jpsiD_IntrinsicCharm} C.H. Chang and W.S. Hou, \Journal{\PRD}{64}{071501}{2001}
%\bibitem{cite:jpsiD_pQCD} G. Eilam, M. Ladisa and Y.D. Yang, \Journal{\PRD}{65}{037504}{2002}.
%\bibitem{cite:jpsiDpi_babar} B. Aubert {\it et al.} [\babar\ collaboration], \verb|hep-ex|/0406022.
\bibitem{cite:jpsiD_hep} B. Aubert {\it et al.} [\babar\ collaboration], \verb|hep-ex|/0503021 and references therein.


%%%%%%%%% Xic
%\bibitem{cite:Xic0_ratio} J.G. K\"orner and M. Kr\"amer, \Journal{\ZPC}{55}{659}{1992}.
%\bibitem{cite:Xic_ccs1} P. Ball and H.G. Dosch, \Journal{\ZPC}{51}{445}{1991}.
%\bibitem{cite:Xic_ccs2} V.L. Chernyak and I.R. Zhitnitsky, \Journal{Nucl. Phys. \textbf{B}}{345}{137}{1990}.
%\bibitem{cite:Xic_ccs3} S.M. Sheikholeslami and M.P. Khanna, \Journal{\PRD}{44}{770}{1991}.
%\bibitem{cite:Xic_ccs4} I. Dunietz {\it et al.}, \Journal{\PRL}{79}{3599}{1997}.
%\bibitem{cite:babar_ccs} B. Aubert {\it et al} [\babar\ collaboration], \Journal{\PRD}{70}{091106}{2004}.
\bibitem{cite:Xic_paper} B. Aubert {\it et al} [\babar\ collaboration], \verb|hep-ex|/0504014 and references therin.
\end{thebibliography}
\end{document}